\documentclass{svmult}

\usepackage{graphicx}
\usepackage{amsmath}
\usepackage{amsfonts}
\usepackage{bm}
\usepackage{bbm}
\usepackage{url}
\usepackage{amssymb}
\usepackage{mathrsfs}
\usepackage{dsfont}
\usepackage{subfigure}
\usepackage{paralist}

\newlength{\irrl}
\newlength{\irrw}

\newcommand{\tr}{\operatorname{tr}}
\newcommand{\curl}{\operatorname{curl}}
\newcommand{\diver}{\operatorname{div}}

\newcommand{\ElE}{\operatorname{E}}
\newcommand{\ElK}{\operatorname{K}}
\newcommand{\ElF}{\operatorname{F}}

\newcommand{\zero}{\bm{0}}

\newcommand{\e}{\bm{e}}
\newcommand{\twist}{\bm{t}}

\newcommand{\n}{\bm{n}}

\newcommand{\N}{\nabla\n}
\newcommand{\free}{\mathscr{F}}
\newcommand{\dens}{f_e}
\newcommand{\field}{\bm{E}}
\newcommand{\elan}{\mathbb{K}(\n)}
\newcommand{\ela}{\mathbb{K}}
\newcommand{\ea}{\varepsilon_\mathrm{a}}
\newcommand{\ez}{\varepsilon_0}

\newcommand{\ground}{\bm{n}_0}
\newcommand{\cohe}{\xi_E}
\newcommand{\cohec}{\xi_E^\mathrm{(c)}}

\newcommand{\body}{\mathscr{B}}
\newcommand{\boundary}{{\partial\body}}

\newcommand{\sphere}{{\mathbb{S}^2}}
\newcommand{\Circle}{{\mathbb{S}^1}}

\newcommand{\ca}{\vartheta}
\newcommand{\cac}{\vartheta_\mathrm{c}}
\newcommand{\az}{\varphi}

\newcommand{\Twist}{\mathbf{T}}

\newcommand{\T}{^\mathsf{T}}

\newcommand{\degree}{^\circ}
\newcommand{\TBN}{$\mathrm{N_{tb}}$\ }

\begin{document}
\title*{Elasticity of Twist-Bend Nematic Phases}
\author{Epifanio G. Virga}
\institute{Epifanio G. Virga \at Dipartimento di Matematica, Universit\`a di Pavia, Via Ferrata 5, I-27100 Pavia, Italy\\ \email{eg.virga@unipv.it}}

\maketitle

\abstract{The ground state of twist-bend nematic  liquid crystals is a \emph{heliconical} molecular arrangement in which the nematic director precesses uniformly about an axis, making a fixed angle with it. Both precession senses are allowed in the ground state of these phases. When one of the two \emph{helicities} is prescribed, a single \emph{helical nematic} phase emerges. A quadratic elastic theory is proposed here for each of these phases which features the same elastic constants as the classical theory of the nematic phase, requiring all of them to be positive. To describe the helix axis, it introduces an extra director field which becomes redundant for ordinary nematics. Putting together helical nematics with opposite helicities, we reconstruct a twist-bend nematic, for which the quadratic elastic energies of the two helical variants are combined in a non-convex energy.}

\keywords{Twist-bend nematic phases; Helical nematic phases;  Freedericks transition; Double-well elastic energies.}

\section{Introduction}\label{sec:intro}
A new liquid crystal equilibrium phase was recently discovered \cite{cestari:phase,chen:chiral,borshch:nematic,chen:twist-bend}. Any such discovery is \emph{per se} a rare event, but this was even more striking as in some specific materials an achiral phase which had long been known was shown to conceal a chiral mutant, attainable upon cooling through a weakly first-order transition. This is known as the twist-bend nematic ($\mathrm{N_{tb}}$) phase.\footnote{The name \emph{twist-bend} was introduced by Dozov~\cite{dozov:spontaneous} together with \emph{splay-bend} to indicate the two alternative nematic distortions which, unlike pure bend and pure splay, can fill the three-dimensional space, as previously observed by Meyer~\cite{meyer:structural}.} Molecular bend seems to be necessary for such a phase to be displayed, but it is not sufficient, as most bent-core molecules fail to exhibit it \cite{jakli:liquid}. In the molecular architecture capable of inducing the $\mathrm{N_{tb}}$ phase, dimers with rigid cores are connected by sufficiently flexible linkers.\footnote{Very recently, experimental evidence has also been provided for \TBN phases arising from \emph{rigid} bent-core molecules \cite{chen:twist-bend}.} The molecular effective curvature, while inducing \emph{no} microscopic \emph{twist}, allegedly favors a \emph{chiral} collective arrangement in which bow-shaped molecules uniformly precess along an ideal cylindrical \emph{helix}.

\begin{figure}
  \centering
  \subfigure[]{\includegraphics[width=.35\linewidth,angle=0]{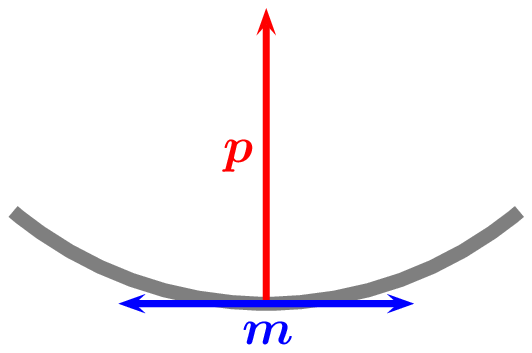}}
  \hspace{.05\linewidth}
  \subfigure[]{\includegraphics[width=.2\linewidth,angle=0]{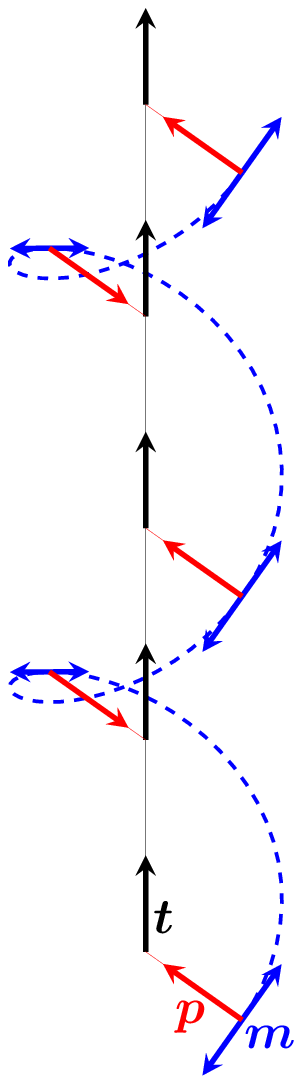}}
  \caption{(a) Molecular achiral model with two symmetry axes, one polar, $\bm{p}$, and one apolar, $\bm{m}$. (b) One variant of the helical nematic phase with helix axis $\twist$. In the other variant, not shown here, the helix winds downwards, in the direction opposite to $\twist$.}
\label{fig:spiral_arc}
\end{figure}

Figure~\ref{fig:spiral_arc} sketches the picture envisaged here.
A single bow-shaped molecule exhibits two symmetry axes, represented by the unit vectors $\bm{p}$ and $\bm{m}$, polar the former and apolar the latter.\footnote{Despite a visual illusion caused by the curved arc, in Fig.~\ref{fig:spiral_arc}(a) the lengths of $\bm{p}$ and $\bm{m}$ are just the same.} The local symmetry point-group is $\mathsf{C}_{2v}$, but, as explained by Lorman and Mettout~\cite{lorman:unconventional,lorman:theory}, by combining the symmetries broken in the helical arrangement in Fig.~\ref{fig:spiral_arc}(b), namely, the continuous translations along the helix axis and the continuous rotations around that axis, a symmetry is recovered which involves any given translation along the helix axis, provided it is accompanied by an appropriately tuned rotation. This forbids any smectic modulation in the mass density, rendering the helical phase purely nematic. While the nematic director $\n$ is defined as the ensemble average $\n:=\langle\bm{m}\rangle$, no polar order survives in a helical phase, as $\langle\bm{p}\rangle=\bm{0}$.\footnote{For this reason, no macroscopic analogue of $\bm{p}$ will be introduced in the theory, and the phase will be treated as macroscopically \emph{apolar}.} Fig.~\ref{fig:spiral_arc}(b) shows only one of the chiral variants that symmetry allows in a \TBN phase; the other winds in the direction opposite to $\twist$. In both cases, $\n$ makes a fixed \emph{cone angle} $\ca$ with $\twist$. For definiteness, we shall call each chiral variant a \emph{helical} nematic phase. In a different language, each helical nematic phase is precisely the $C$-phase predicted by Lorman and Mettout \cite{lorman:unconventional,lorman:theory}, which breaks spontaneously the molecular chiral symmetry, producing two equivalent macroscopic variants with opposite chiralities (see Fig.~\ref{fig:spiral_arc}). This paper is intended to study separately each helical nematic variant hosted in a \TBN phase. How opposite variants may be brought in contact within a purely elastic theory will be the subject of a forthcoming paper \cite{virga:double-well}.

Though, in retrospect, a number of experimental studies had already anticipated \TBN phases (see, for example, \cite{sepelj:intercalated,imrie:liquid,panov:spontaneous}, to cite just a few), a clear identification of the new phase was achieved in \cite{cestari:phase} by a combination of methods (see also \cite{henderson:methylene,cestari:crucial,panov:microsecond,panov:field-induced,beguin:chirality}), and an impressive direct evidence for it has been provided in \cite{chen:chiral,borshch:nematic,chen:twist-bend} (see also \cite{copic:nematic}), with accurate measurements of both the helix pitch ($\approx10\,\mathrm{nm}$) and cone angle ($\approx20\degree$).
Theoretically, $\mathrm{N_{tb}}$ phases had already been predicted by Meyer~\cite{meyer:structural} and Dozov~\cite{dozov:spontaneous}, from two different perspectives, the former inspired by the symmetry of polar electrostatic interactions (a line of thought recently further pursued in \cite{shamid:statistical}) and the latter starting from purely elastic (and steric) considerations. A helical molecular arrangement was also seen in the molecular simulations of Memmer~\cite{memmer:liquid}, who considered bent-core Gay-Berne molecules with \emph{no} polar electrostatic interactions, though featuring an effective shape polarity.

Dozov's $\mathrm{N_{tb}}$ theory requires a \emph{negative} bend elastic constant, which is compatible with the boundedness (from below) of the total energy only if appropriate \emph{quartic} corrections are introduced in the energy density. However, a large number of these terms are allowed by symmetry \cite{dozov:spontaneous}, and the theory may realistically be applied only by choosing just a few of them and neglecting all the others \cite{meyer:flexoelectrically}. Moreover, recent experiments \cite{adlem:chemically} have reported an increase in the (positive) bend constant measured in the nematic phase near the transition to the $\mathrm{N_{tb}}$ phase. In an attempt to justify the negative elastic bend constant required by Dozov's theory, a recent study has replaced it with an \emph{effective} bend constant resulting from the coupling with the polarization characteristic of flexo-elasticity \cite{shamid:statistical}. As a consequence, however, the twist-bend modulated phase envisaged in \cite{shamid:statistical} is locally \emph{ferroelectric}, whereas, as explained by the symmetry argument of Lorman and Mettout~\cite{lorman:theory,lorman:unconventional}, each helical nematic phase is expected to be \emph{apolar}.

The theories in \cite{dozov:spontaneous} and \cite{shamid:statistical} are not in contradiction with one another, the only difference being that the latter justifies a negative bend elastic constant as commanded by the very bend-polarization coupling that gives rise to a modulated polar phase. There is still a conceptual difference between these theories: the former is purely elastic but quartic, whereas the latter is quadratic but flexo-electric. Being highly localized, the ferroelectricity envisaged in \cite{shamid:statistical} does not produce any average macroscopic polarization, and so it would be compatible with the current experimental observations which have not found so far any trace of macroscopic ferroelectricity. This, however, leaves the question unanswered as to whether an intrinsically quadratic, purely elastic theory could also explain \TBN phases.
In this work, we propose such a theory; it will feature the same elastic constants as Frank's classical theory \cite{frank:theory}, with a positive bend constant and an extra director field. This theory reduces to Frank's for ordinary nematics, for which the extra director becomes redundant.

This paper is organized as follows. In Section~\ref{sec:helical}, we shall formally introduce a helical nematic phase, defined as each \TBN variant with a prescribed \emph{helicity} in its ground state.\footnote{See \eqref{eq:helicity} for a precise definition.} A quadratic elastic free energy will be considered for each helical nematic variant, under the (temporary) assumption that they can be thought of as separate manifestations of one and the same \TBN phase. For a given sign of the helicity, negative for definiteness, we shall apply the proposed elastic theory to two classical instabilities, namely, the helix unwinding first encountered in chiral nematics (in Section~\ref{sec:helix}), and the Freedericks transition that has long made it possible to measure the elastic constants of ordinary nematics (in Section~\ref{sec:freedericks}). Both these applications will acquire some new nuances within the present theory. In Section~\ref{sec:Double-Well}, we first derive the quadratic elastic energy density for a helical nematic phase with \emph{positive} helicity. The helical nematic variants with both helicities are then combined together in a \TBN phase; the corresponding elastic energy density need to attain one and the same minimum in two separate \emph{wells}. There are several ways to construct such an energy, which by necessity will \emph{not} be \emph{convex}; we shall build upon the quadratic elastic energy for a single helical phase arrived at in Section~\ref{sec:helical}. In the two ways that we consider in detail, the elastic energy density for a \TBN phase features only four elastic constants. Finally, in Section~\ref{sec:conclusion}, we collect the conclusions reached in this work and comment on some possible avenues for future research.

\section{Helical nematic phases}\label{sec:helical}
A \TBN differs from classical nematics in its ground state, the state relative to which the elastic cost of all distortions is to be accounted for. The ground state of a classical nematic is the uniform alignment (in an arbitrary direction) of the nematic director $\n$. The $\mathrm{N_{tb}}$ ground state is a \emph{heliconical} twist, in which $\n$ performs a uniform precession, making everywhere the angle $\ca$ with a helix axis, $\twist$, arbitrarily oriented in space.
Such a ground state should reflect the intrinsically less distorted molecular arrangement that results from minimizing the interaction energy of the achiral, bent molecules that comprise the medium.

By symmetry, there are indeed \emph{two} such states, distinguished by the sense of precession (either clockwise or anticlockwise around $\twist$). In general, borrowing a definition from Fluid Dynamics (see, for example, \cite{moffatt:degree}), we call the pseudoscalar
\begin{equation}\label{eq:helicity}
\eta:=\n\cdot\curl\n,
\end{equation}
the \emph{helicity} of the director field $\n$. We shall see now that it is precisely the sign of $\eta$ that differentiates the two variants of the ground state of a $\mathrm{N_{tb}}$.

\begin{figure}
\centering
\includegraphics[width=.5\linewidth,angle=0]{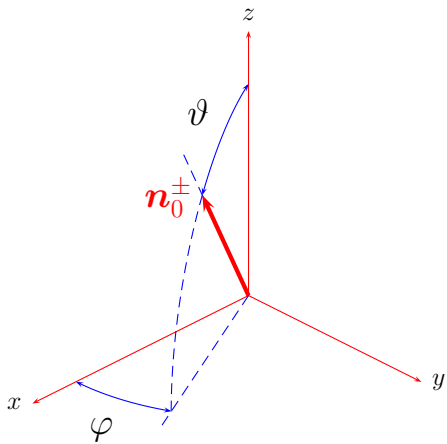}
\caption{The angles $\ca$ and $\az=\pm qz+\az_0$ that in \eqref{eq:ground} represent the fields $\ground^\pm$ are illustrated in a Cartesian frame $(x,y,z)$. Both $\ca$ and $q$ are fixed parameters characteristic of the phase.}
\label{fig:spherical}
\end{figure}

Letting $\twist$ coincide with the unit vector $\e_z$ of a Cartesian frame $(x,y,z)$, the fields $\ground^\pm$ representing the ground states can be written in the form
\begin{equation}\label{eq:ground}
\begin{split}
\ground^\pm=&\sin\ca\cos(\pm qz+\az_0)\,\e_x\\+&\sin\ca\sin(\pm qz+\az_0)\,\e_y+\cos\ca\,\e_z,
\end{split}
\end{equation}
where $\az_0$ is an arbitrary phase angle, $q$ is a prescribed \emph{wave parameter}, taken to be non-negative, as characteristic of the condensed phase as the \emph{cone angle} $\ca$ (see Fig.~\ref{fig:spherical}).
The \emph{pitch} $p$ corresponding to $q$ is given by\footnote{No confusion should arise here between the pitch $p$ of the \TBN phase and the polar vector $\bm{p}$ mentioned in the Introduction. The former is macroscopic in nature, whereas the latter is microscopic.}
\begin{equation*}\label{eq:p}
p:=\frac{2\pi}{q}.
\end{equation*}
On every two planes orthogonal to $\e_z$ and $p$ apart, each field $\ground^\pm$ in \eqref{eq:ground} delivers one and the same nematic director. A simple computation shows that
\begin{equation}\label{eq:helicity_ground_state}
\eta^\pm:=\ground^\pm\cdot\curl\ground^\pm=\mp q\sin^2\ca.
\end{equation}

We shall call \emph{helical nematic} each of the two phases for which either $\ground^+$ or $\ground^-$ is the nematic field representing the ground state. Here, for definiteness, we shall develop our theory as if only $\ground^+$ represented the ground state. By \eqref{eq:helicity_ground_state}, such a state has \emph{negative} helicity. The corresponding case of positive helicity will be studied in Section~\ref{sec:positive}. Until then we shall drop the superscript $^+$ from $\ground^+$ to avoid clatter.

\subsection{Negative helicity}\label{sec:negative}
It readily follows from \eqref{eq:ground} that\footnote{Here and below $\times$ and $\otimes$ denote the vector and tensor products of vectors. In particular, for any two vectors, $\bm{a}$ and $\bm{b}$, $\bm{a}\otimes\bm{b}$ is the second-rank tensor that acts as follows on a generic vector $\bm{v}$, $(\bm{a}\otimes\bm{b})\bm{v}:=(\bm{b}\cdot\bm{v})\bm{a}$, where $\cdot$ denotes the inner product of vectors. An alternative way of denoting the dyadic product $\bm{a}\otimes\bm{b}$ would be simply $\bm{a}\bm{b}$. In Cartesian components, they are both represented as $a_ib_j$.}
\begin{equation}\label{eq:grad_ground}
\nabla\ground=q\left(\e_z\times\ground\right)\otimes\e_z.
\end{equation}
More generally, for $\n$ prescribed at a point in space, the tensor
\begin{equation}\label{eq:natural_distortion}
\Twist:=q(\twist\times\n)\otimes\twist
\end{equation}
expresses the \emph{natural} distortion\footnote{A natural distortion is a distortion present in the ground state, when the latter fails to be the uniform nematic field $\n$ for which $\N\equiv\zero$.} that would be associated at that point with the preferred helical configuration agreeing with the prescribed nematic director $\n$ and having $\twist$ as helix axis. We imagine that in the absence of any frustrating cause, given $\n$ at a point, the director field would attain in its vicinity a spatial arrangement such that
\begin{equation}\label{eq:local_and_natural_distortions}
\nabla\n=\Twist,
\end{equation}
where $\Twist$ is as in \eqref{eq:natural_distortion} and $\twist$ is any unit vector such that
\begin{equation}\label{eq:cone_constraint}
\n\cdot\twist=\cos\ca.
\end{equation}
This would make \eqref{eq:grad_ground} locally satisfied, even though $\n$ does not coincide with $\ground$ in the large. Correspondingly, the elastic energy that would locally measure the distortional cost should be measured relative to the whole class of natural distortions, vanishing whenever any of the latter is attained. With this in mind, we write the elastic energy density $f^-_e$ in the form\footnote{The superscript $^-$ will remind us that the ground state of the helical nematic phase we are considering here has a negative helicity $\eta$,}
\begin{equation}\label{eq:elastic_energy_density_definition}
f^-_e(\twist,\n,\N)=\frac12(\N-\Twist)\cdot\elan[\N-\Twist],
\end{equation}
where $\elan$ is the most general positive-definite, symmetric fourth-order tensor invariant under rotations about $\n$. Clearly, if for given $\n$ and $\N$, $\twist$ can be chosen in \eqref{eq:elastic_energy_density_definition} so that \eqref{eq:local_and_natural_distortions} is satisfied, $f^-_e$ vanishes, attaining its absolute minimum. On the contrary, if  there is no such $\twist$, then minimizing $f^-_e$ in $\twist$ would identify the natural, undistorted state closest to the nematic distortion represented by $\N$ in the metric induced by $\elan$. For this reason, here both $\n$ and $\twist$ are to be considered as unknown fields linked by \eqref{eq:cone_constraint}: at equilibrium, the free-energy functional that we shall construct is to be minimized in both these fields.

Combining the general representation formula for $\elan$ and the identities\footnote{The superscript $\T$ means tensor transposition.}
\begin{equation*}\label{eq:tensor_identities}
(\N)\T\n=\zero,\qquad\Twist\T\n=\zero,
\end{equation*}
we can reduce $\ela$ in \eqref{eq:elastic_energy_density_definition} to the following equivalent form
\begin{equation}\label{eq:K_reduced_representation}
\ela_{ijhk}=k_1\delta_{ih}\delta_{jk}+k_2\delta_{ij}\delta_{hk}+k_3\delta_{ih}n_jn_k
+k_4\delta_{ik}\delta_{jh},
\end{equation}
where $k_1$, $k_2$, $k_3$, and $k_4$ are elastic moduli (as in \cite[p.\,114]{virga:variational}). By use of \eqref{eq:K_reduced_representation}, of \eqref{eq:cone_constraint}, and
\begin{equation*}\label{eq:scalar_identities}
\tr\Twist=0,\qquad \N\cdot\Twist=q(\twist\times\n)\cdot(\N)\twist,
\end{equation*}
we transform \eqref{eq:elastic_energy_density_definition} into
\begin{equation}\label{eq:elastic_energy_density_transformed}
\begin{split}
f^-_e(\twist,\n,\N)
&=\frac12\Bigl\{K_{11}(\diver\n)^2+K_{22}(\n\cdot\curl\n+q|\twist\times\n|^2)^2\\
&+K_{33}|\n\times\curl\n+q(\n\cdot\twist)\,\twist\times\n|^2\\
&+K_{24}[\tr(\N)^2-(\diver\n)^2]\Bigr\}
-K_{24}q\,\twist\times\n\cdot(\N)\T\twist,
\end{split}
\end{equation}
where $K_{11}$, $K_{22}$, $K_{33}$, and $K_{24}$, which are analogous to the classical Frank's constants \cite{frank:theory}, are related to the moduli $k_1$, $k_2$, $k_3$, and $k_4$ through the equations
\begin{equation*}\label{eq:K_and_k}
K_{11}=k_1+k_2+k_4,\quad K_{22}=k_1,\quad
K_{33}=k_1+k_3,\quad K_{24}=k_1+k_4.
\end{equation*}
For $f^-_e$ in \eqref{eq:elastic_energy_density_transformed} to be positive definite, the elastic constants $K_{11}$, $K_{22}$, $K_{33}$, and $K_{24}$ must obey the inequalities,
\begin{equation*}\label{eq:Ericksen_inequalities}
2K_{11}\geqq K_{24},\quad 2K_{22}\geqq K_{24},\quad
K_{33}\geqq0,\quad K_{24}\geqq0,
\end{equation*}
which coincide with the classical Ericksen's inequalities for ordinary nematics \cite{ericksen:inequalities}.

The total elastic energy $\free^-_e$ is represented by the functional
\begin{equation*}\label{eq:elastic_energy_functional}
\free^-_e[\twist,\n]:=\int_\body f^-_e(\twist,\n,\N)dV,
\end{equation*}
where both $\twist$ and $\n$ are subject to the pointwise constraint \eqref{eq:cone_constraint} and either of them is prescribed on the boundary $\boundary$ of the region in space occupied by the medium.
It is worth recalling that both $\twist$ and $\n$ are fields that need to be determined so as to obey \eqref{eq:cone_constraint} and to minimize $\free^-_e$. In this theory, the helix axis of the preferred conical state and the nematic director of the actual molecular organization equally participate in the energy minimization with the objective of reducing the discrepancy between natural and actual nematic distortions.
Physically, $\twist$ represents the optic axis of the medium, likely to be the only optic observable when the pitch $p$ ranges in the the nanometric domain.

Such an abundance of state variables is a direct consequence of the degeneracy in the ground state admitted for helical nematics. The \emph{vacuum manifold}, as is often called the set of distortions that minimize $f^-_e$, is indeed a three-dimensional orbit of congruent cones, identifiable with $\sphere\times\Circle$, where $\sphere$ is the unit sphere and $\Circle$ the unit circle in three space dimensions. By \eqref{eq:natural_distortion}, all natural distortions $\Twist$ are represented by $\twist\in\sphere$ and any $\n$ in the cone of semi-amplitude $\ca$ around $\twist$. This shows, yet in another language, how rich in states is the single well where $f^-_e$ vanishes.

A number of remarks are suggested by formula \eqref{eq:elastic_energy_density_transformed}. First, it reduces to the elastic free-energy density of ordinary nematics, which features only $\n$, when either the wave parameter $q$ or the cone angle $\ca$ vanish, thus indicating two possible mechanisms to induce helicity in an ordinary nematic. Second, for $\ca=\frac\pi2$, it delivers an alternative energy density for chiral nematics, which is positive-definite for all $K_{24}\geqq0$ (whereas, to ensure energy positive-definiteness, the classical theory requires that $K_{24}=0$, see \cite[p.\,127]{virga:variational}). Finally, for arbitrary $q>0$ and $0<\ca<\frac{\pi}{2}$, $f^-_e$ in \eqref{eq:elastic_energy_density_transformed} is invariant under the reversal of either $\twist$ or $\n$, showing that both fields enjoy the nematic symmetry.

\section{Helix unwinding}\label{sec:helix}
In the presence of an external field, say an electric field $\field$, the free-energy density acquires an extra term, which we write as
\begin{equation*}\label{eq:electric_energy density}
f_E(\n)=-\frac12\ez\ea E^2(\n\cdot\e)^2,
\end{equation*}
where $\ez$ is the vacuum permittivity, $\ea$ is the (relative) dielectric anisotropy, and we have set $\field=E\e$, with $E>0$ and $\e$ a unit vector. Accordingly, the total free-energy functional $\free^-$ is defined as
\begin{equation}\label{eq:free_energy_total}
\free^-[\twist,\n]:=\int_\body\{f^-_e(\twist,\n,\N)+f_E(\n)\}dV.
\end{equation}

To minimize $\free^-$ when $\body$ is the whole space, we shall assume that $\twist$ is uniform and $\n$ is spatially periodic with period $L$ and we shall compute the average $F^-$ of $\free^-$ over an infinite slab of thickness $L$ orthogonal to $\twist$. Letting $\twist$ coincide with the unit vector $\e_z$ of a given Cartesian frame $(x,y,z)$, we represent $\n$ in precisely the same form adopted in \eqref{eq:ground} for $\ground$, but with $qz+\az_0$ replaced by a function $\az=\az(z)$ such that
\begin{equation*}\label{eq:periodicity_azimut}
\az(0)=0\quad\text{and}\quad\az(L)=2\pi.
\end{equation*}
With no loss of generality, we may choose $\e$ in the $(y,z)$ plane and represent it as
\begin{equation*}\label{eq:a_representation}
\e=\cos\psi\,\e_z+\sin\psi\,\e_y.
\end{equation*}
Computing $F^-$ on the ground state $\az=2\pi z/p$ (where $L=p$), one easily sees that the average energy is smaller for $\psi=\frac\pi2$ than for $\psi=0$, whenever either
\begin{enumerate}[(a)]
\item\label{case:1} $\ea<0$ and $\ca<\cac:=\arctan\sqrt{2}\doteq54.7\degree$, or
\item\label{case:2} $\ea>0$ and $\ca>\cac$,
\end{enumerate}
which are the only cases we shall consider here. In case \eqref{case:1} (and for $\psi=\frac\pi2$), $F^-$ reduces to
\begin{equation*}\label{eq:F_unwinding}
\begin{split}
F^-[L,\az]=K\sin^2\ca\left\{\frac{1}{2L}\int_0^L\left[\az'^2+\frac{1}{\cohe^2}\sin^2\az\right]dz -\frac{2\pi q}{L} \right\},
\end{split}
\end{equation*}
where
\begin{equation*}\label{eq:K_effective}
K:=K_{22}\sin^2\ca+K_{33}\cos^2\ca
\end{equation*}
is an effective \emph{twist-bend} elastic constant and
\begin{equation}\label{eq:coherence_length}
\cohe:=\frac{1}{E}\sqrt{\frac{K}{\ez|\ea|}}
\end{equation}
is a field \emph{coherence length}.

Minimizing $F^-$ in both $L$\ and $\az$ is a problem formally akin to the classical problem of unwinding the cholesteric helix \cite{degennes:calul,meyer:effects,meyer:distortion}. The minimizing $\az$ is determined implicitly by
\begin{equation*}\label{eq:phi_unwinding}
\frac{z}{L}=\frac14\left(\frac{\ElF(\az+\frac\pi2,k)}{\ElK(k)}-1\right),
\end{equation*}
where $\ElF$ and $\ElK$ are the elliptic and complete elliptic integrals of the first kind, respectively, and $k$ is the root in the interval $[0,1]$ of the equation
\begin{equation}\label{eq:root_k}
\frac{\ElE(k)}{k}=\pi^2\frac{\cohe}{p},
\end{equation}
where $\ElE$ is the complete elliptic integral of the second kind \cite[p.\,486]{olver:NHMF}. Equation \eqref{eq:root_k}
has a (unique) root only for
\begin{equation*}
\cohe\geqq\cohec:=\frac{p}{\pi^2},
\end{equation*}
for which $L$ is correspondingly delivered by
\begin{equation*}\label{eq:root_L}
L=4\cohe k\ElK(k).
\end{equation*}
\begin{figure}[h]
\centering
\includegraphics[width=0.7\linewidth]{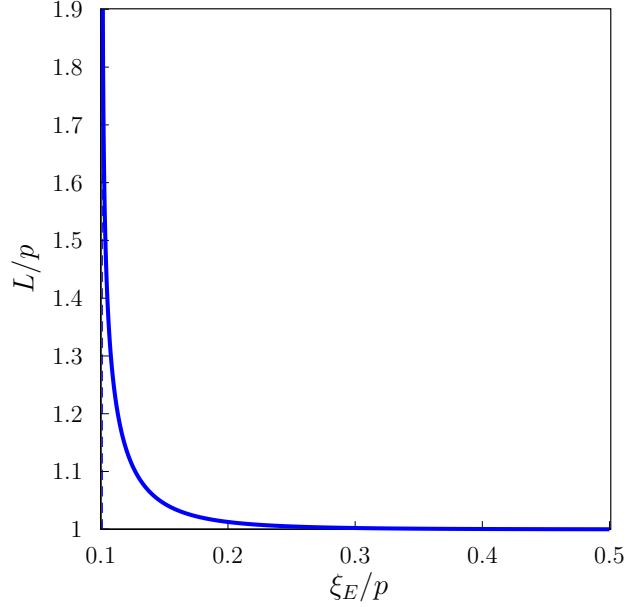}
\caption{
The graph of the spatial period $L$ of the distorted helix under field (scaled to the undistorted pitch $p$) against the field coherence length $\cohe$ in \eqref{eq:coherence_length}
(equally scaled to $p$). The complete unwinding takes place for $\cohe/p=1/\pi^2\doteq0.101$, where the graph diverges.}
\label{fig:L_graph}
\end{figure}
Figure~\ref{fig:L_graph} shows the graph of $L$ against $\cohe$, which diverges as $\cohe$ approaches $\cohec$. As for ordinary chiral nematics, a field with coherence length $\cohec$ unwinds completely the helix of a $\mathrm{N_{tb}}$, but its actual strength now depends on the elastic constants $K_{22}$ and $K_{33}$, and the cone angle $\ca$.
The measured values of $p$ range in the domain of tens of nanometers. Thus, the steepness of the graph in Fig.~\ref{fig:L_graph} along with \eqref{eq:coherence_length} suggest that in actual terms the field should be very strong for any unwinding to be noticed.\footnote{A simple estimate based on \eqref{eq:root_k} would show that a field strength larger than $100\,\mathrm{V}/\mu\mathrm{m}$ is needed to unwind the \TBN helix even if we assume $\ea$ as large as $10$ and $K$ as small as $1\,\mathrm{pN}$.}

\section{Freedericks transition}\label{sec:freedericks}
In ordinary nematics, the Freedericks transition is an instability that enables one to measure the classical Frank's elastic constants. For a $\mathrm{N_{tb}}$, the setting is complicated by the role played by the additional field $\twist$.

We consider a $\mathrm{N_{tb}}$ liquid crystal confined between two parallel plates, placed in a Cartesian frame $(x,y,z)$ at $z=0$ and $z=d$, respectively. On both plates, we subject $\n$ to a \emph{conical} anchoring with respect to the plates' normal $\e_z$ at precisely the cone angle $\ca$, so that, by the constraint \eqref{eq:cone_constraint}, $\twist$ is there prescribed to coincide with $\e_z$. Within the infinite cell bounded by these plates we allow $\twist$ to vary in the $(y,z)$ plane, so that
\begin{equation*}\label{eq:FrederikS_e_representation}
\twist=\cos\psi\,\e_z+\sin\psi\,\e_y,
\end{equation*}
where $\psi=\psi(z)$. Letting
\begin{equation*}
\twist_\perp:=\sin\psi\,\e_z-\cos\psi\,\e_y,
\end{equation*}
we represent the nematic director as
\begin{equation}\label{eq:Freedericks_n_representation}
\n=\cos\ca\,\twist+\sin\ca\cos\az\,\twist_\perp+\sin\ca\sin\az\,\e_x,
\end{equation}
where $\az=\az(z)$ is the precession angle. The function $\psi$ is subject to the conditions
\begin{equation}\label{eq:Freedericks_psi_boundary_conditions}
\psi(0)=\psi(d)=0,
\end{equation}
while $\az$ is free in the whole of $[0,d]$. The system is further subjected to an external field, $\field=E\e_z$. In the following analysis, we shall assume that either of the aforementioned  cases \eqref{case:1} or \eqref{case:2} occurs here.

Taking $p/d\ll1$ and treating $\psi$ as a perturbation, of which we only retain quadratic terms in the energy, when the precession angle represents the undistorted helix, $\az=2\pi z/p$, we obtain the average $F^-$ of the free-energy functional $\free^-$ in \eqref{eq:free_energy_total}
(per unit area of the plates) as
\begin{equation}\label{eq:Freedericks_average_free_energy}
F^-[\psi]=\frac12 K_{33}\int_0^d\left\{A\psi'^2+Bq^2\psi^2\right\}dz,
\end{equation}
where
\begin{gather*}
A:=\frac12\Big[\sin^2\ca(k_{11}+k_{22}\cos^2\ca)+\cos^2\ca(1+\cos^2\ca)\Big],
\\
B:=\frac12\Big[\sin^2\ca(k_{11}+k_{22}\cos^2\ca+\sin^2\ca)
-\frac{1}{(\cohe q)^2}|2\cos^2\ca-\sin^2\ca|\Big],
\\
k_{11}:=\frac{K_{11}}{K_{33}},\quad k_{22}:=\frac{K_{22}}{K_{33}},\quad\cohe:=\frac{1}{E}\sqrt{\frac{K_{33}}{\ez|\ea|}}.
\end{gather*}
It is now easily seen that $\psi\equiv0$ is a locally stable extremum of the functional $F^-$
in \eqref{eq:Freedericks_average_free_energy} subject to \eqref{eq:Freedericks_psi_boundary_conditions}
for $\cohe>\cohec$, whereas it is locally unstable for $\cohe<\cohec$, where
\begin{equation}\label{eq:Freedericks_critical_xi_E}
\frac{\cohec}{p}:=\frac{\sqrt{|2-t^2|(1+t^2)}}{\pi\sqrt{4[(1+k_{11})t^2+k_{11}+k_{22}]t^2+ \frac{p^2}{d^2}[k_{11}t^4+(1+k_{11}+k_{22})t^2+2]}}
\end{equation}
and $t:=\tan\ca$. Both this and the helix unwinding treated in Section~\ref{sec:helix} are continuous transitions.

Since in this theory $p/d$ is small, $\cohec$ is only weakly dependent on the cell thickness $d$. It should be noted however that for $\ca=0$, \eqref{eq:Freedericks_critical_xi_E} reduces to $\cohec=d/\pi$, which reproduces the classical Freedericks threshold \cite[p.\,179]{virga:variational}.

In the special, hypothetical case of equal elastic constants (for which $k_{11}=k_{22}=1$), $\cohec$ acquires a simpler form, which retains the qualitative features of \eqref{eq:Freedericks_critical_xi_E}:
\begin{equation}\label{eq:Freedericks_critical_xi_E_equal_constants}
\frac{\cohec}{p}=\frac{\sqrt{|2-t^2|}}{\pi\sqrt{8t^2+\frac{p^2}{d^2}\frac{t^4+3t^2+2}{1+t^2}}}.
\end{equation}
\begin{figure}
\centering
\includegraphics[width=.9\linewidth]{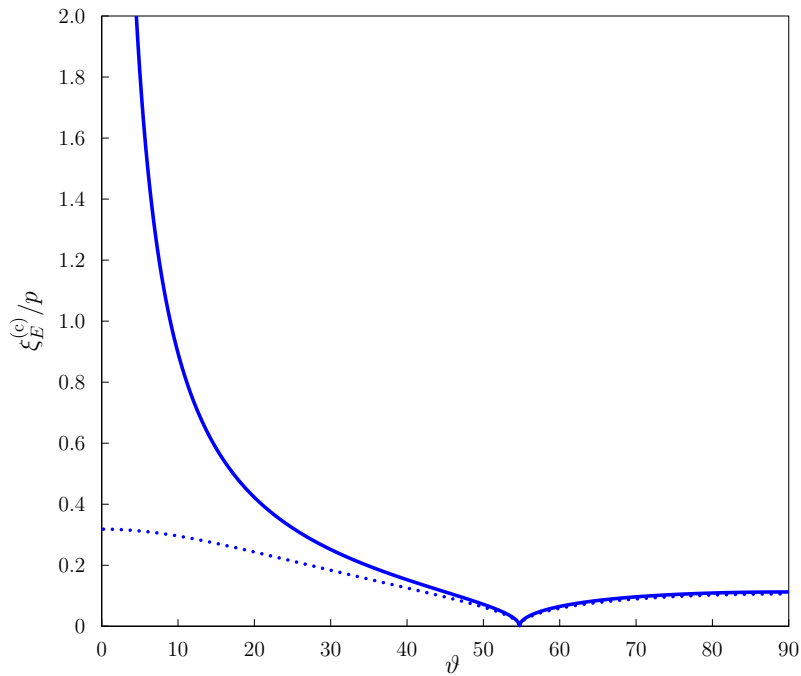}
\caption{
The graphs of $\cohec$ (scaled to the pitch $p$ of the undistorted helix) against the cone angle $\ca$ (expressed in degrees), as delivered by \eqref{eq:Freedericks_critical_xi_E_equal_constants} under the assumption of equal elastic constants, for $p/d=0$ (solid line) and $p/d=1$ (dotted line). For $\ca=0$, the former graph diverges, while the latter reaches the value $\cohec/p=1/\pi\doteq0.318$. For $0<p/d<1$, the graphs of $\cohec$, none of which diverges, fill the region bounded by the graphs shown here.}
\label{fig:F_Graph}
\end{figure}
The graph of $\cohec$
as delivered by \eqref{eq:Freedericks_critical_xi_E_equal_constants}
is plotted in Fig.~\ref{fig:F_Graph} for $p/d=0$ and $p/d=1$, the graphs for all intermediate values of $p/d$ being sandwiched between them. These graphs show that for $\ca$ sufficiently large, in particular for $\ca>\cac$ (and $\ea>0$), $\cohec$ is virtually independent of $d$, whereas it is not so for $\ca$ small. Moreover, for moderate values of $\ca$ and $d$ much larger than $p$, $\cohec$ can easily be made equal to several times the pitch of the undistorted helix.
If, as contemplated by \eqref{eq:Freedericks_critical_xi_E}, the actual field required to ignite the Freedericks transition is weakly dependent on the cell thickness $d$, the corresponding critical potential $U_\mathrm{c}$ would scale almost linearly with $d$.\footnote{In the experiments performed in \cite{borshch:nematic}, it was found that $U_\mathrm{c}\varpropto\sqrt{d}$, but the boundary conditions imposed there seem to differ from the conical boundary conditions considered here. Moreover, in \cite{borshch:nematic} the transition nucleated locally from the inside of the cell instead of happening uniformly, as presumed here. This might suggest that in the real experiment both helical variants present in a \TBN are participating in the transition. It would then be advisable taking with a grain of salt any direct comparison of our theory for helical nematics with experiments available for the whole \TBN phase.}

\section{Double-well energy}\label{sec:Double-Well}
A \TBN phase can be regarded as a mixture of two helical nematic phases with opposite helicities. The ground  states of these phases corresponding to all admissible natural distortions are the members of two symmteric energy wells. In a way, this is reminiscent of the mixture of martensite twins in some solid crystals, which are equi-energetic variants with symmetrically sheared lattices (see, for example, \cite[p.\,129]{muller:entropy}). A thorough mathematical theory of these solid phases is based on a non-convex energy functional in the elastic deformation, featuring a multiplicity of energy wells \cite{ball:fine,ball:proposed}. Below, adapting these ideas to the new context envisaged here, in which the energy depends on the local distortion of the molecular arrangement, and no deformation from a reference configuration is involved, we show how to construct a double-well elastic energy density $\dens$ for a \TBN phase starting from the energy densities for helical nematic phases of opposite helicities. To this end we need first supplement $\dens^-$ in \eqref{eq:elastic_energy_density_transformed} with the appropriate energy density $\dens^+$ for a helical nematic phase of positive helicity.

\subsection{Positive helicity}\label{sec:positive}
Prescribing the helicity of the ground state of a helical nematic phase to be positive, instead of negative as above, would amount to replacing \eqref{eq:elastic_energy_density_definition} by
\begin{equation}\label{eq:elastic_energy_density_definition_positive}
f^+_e(\twist,\n,\N)=\frac12(\N+\Twist)\cdot\elan[\N+\Twist],
\end{equation}
still with $\Twist$ as in \eqref{eq:natural_distortion} and $q>0$. Our development following \eqref{eq:elastic_energy_density_definition} could be repeated verbatim here and it would lead us to the same conclusion obtained by subjecting \eqref{eq:elastic_energy_density_transformed} to the \emph{formal} change of $q$ into $-q$:
\begin{equation}\label{eq:elastic_energy_density_transformed_positive}
\begin{split}
f^+_e(\twist,\n,\N)
&=\frac12\Bigl\{K_{11}(\diver\n)^2+K_{22}(\n\cdot\curl\n-q|\twist\times\n|^2)^2\\
&+K_{33}|\n\times\curl\n-q(\n\cdot\twist)\,\twist\times\n|^2\\
&+K_{24}[\tr(\N)^2-(\diver\n)^2]\Bigr\}
+K_{24}q\,\twist\times\n\cdot(\N)\T\twist,
\end{split}
\end{equation}
where $q$ remains a \emph{positive} parameter.

Clearly, the energy well of \eqref{eq:elastic_energy_density_definition_positive} and \eqref{eq:elastic_energy_density_transformed_positive} is formally obtained from the corresponding well of \eqref{eq:elastic_energy_density_definition} and \eqref{eq:elastic_energy_density_transformed} by a sign inversion. 

\subsection{\TBN free energy density}\label{sec:TBN}
\begin{figure}
  \centering
  \includegraphics[width=.7\linewidth,angle=0]{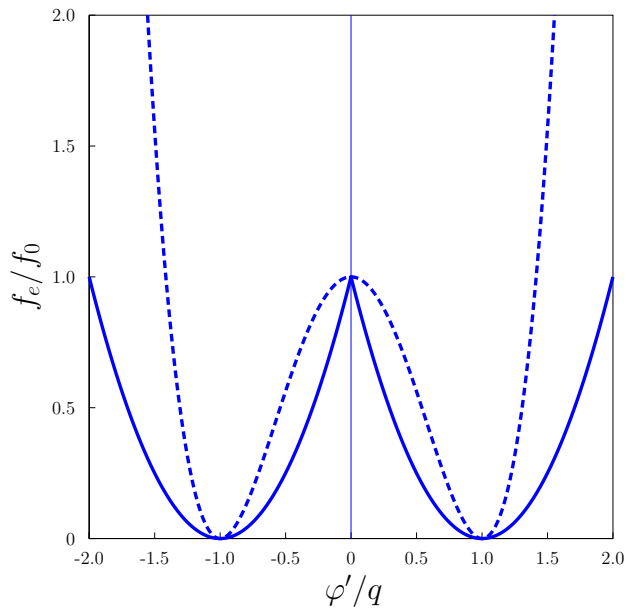}
  \caption{One-dimensional pictures for $\dens$ in \eqref{eq:f_e_TBN_quadratic} (solid line) and \eqref{eq:f_e_TBN_quartic} (dashed line). Here $\az=\az(z)$ would represent the precession angle in a molecular arrangement such as that described by \eqref{eq:Freedericks_n_representation} and $\az'$ is its spatial derivative. Each minimum is representative for a three-dimensional well described by \eqref{eq:natural_distortion} and its mirror image (with $q$ replaced by $-q$).}
\label{fig:Free_Plots}
\end{figure}

Following \cite{truskinovsky:ericksen}, which studied systematically how to extend the non-convex energy first proposed by Ericksen~\cite{ericksen:equilibrium} for a one-dimensional elastic bar, we consider two possible choices for the elastic energy density $\dens$ of a \TBN phase:
\begin{subequations}\label{eq:f_e_TBN}
\begin{equation}\label{eq:f_e_TBN_quadratic}
f_e(\twist,\n,\N)=\min\{f_e^+(\twist,\n,\N),f_e^-(\twist,\n,\N)\},
\end{equation}
\begin{equation}\label{eq:f_e_TBN_quartic}
f_e(\twist,\n,\N)=\frac{1}{f_0}f_e^+(\twist,\n,\N)f_e^-(\twist,\n,\N),
\end{equation}
where
\begin{equation*}
f_0:=\frac12\sin^2\ca(K_{22}\sin^2\ca+ K_{33}\cos^2\ca)
\end{equation*}
is a normalization constant chosen so as to ensure that $\dens(\twist,\n,\zero)=\dens^\pm(\twist,\n,\zero)$.
\end{subequations}
While $\dens$ in \eqref{eq:f_e_TBN_quadratic} is quadratic around each well, it fails to be smooth for $\N=\zero$. On the other hand, $\dens$ in \eqref{eq:f_e_TBN_quartic} is everywhere smooth, but it is quartic. The differences between these energy densities are illustrated pictorially in Fig.~\ref{fig:Free_Plots}.

We shall not explore other possible forms $\dens$. We only heed that both \eqref{eq:f_e_TBN_quadratic} and \eqref{eq:f_e_TBN_quartic} inherit the simple structure of the elastic energy density of a helical nematic phase, which features only four positive elastic constants, as in the classical Frank's theory of ordinary nematics. General considerations on how to match ground states extracted from two different wells of $\dens$ (at zero energy cost) are independent of the peculiar form assumed for this function, as they are only consequences of the structure of each well. A study of the geometric compatibility conditions that arise in this case will be presented elsewhere \cite{virga:double-well}.

\section{Conclusion}\label{sec:conclusion}
The elastic energy density proposed in \eqref{eq:elastic_energy_density_transformed} and \eqref{eq:elastic_energy_density_transformed_positive} to describe the equilibrium distortions of each helical variant of a \TBN phase featured just the classical four elastic constants and introduced the helix axis $\twist$ in addition to the nematic director $\n$. The instabilities studied above illustrated two second-order transitions that differ also qualitatively from their classical analogues. Only experiments may decide at this stage whether a quadratic elastic theory based on either \eqref{eq:f_e_TBN_quadratic} or \eqref{eq:f_e_TBN_quartic} above\footnote{Even if globally quartic, the energy density in \eqref{eq:f_e_TBN_quartic} is quadratic about each energy well. Moreover, it features only the four elastic constants required by the quadratic energy density of a helical phase with prescribed helicity.} is better suited to describe the novel \TBN phases than the quartic theory proposed in \cite{dozov:spontaneous}. The instabilities described in this paper for each \TBN variant  just provide a theoretical means to set the quadratic theory to the test.

Two director fields, namely, $\n$ and $\twist$, were deemed necessary here to describe the local distortion of a \TBN phase.
This poses the question as to which defects both fields may exhibit and how they are interwoven, in view of the constraint \eqref{eq:cone_constraint}.
An extra field also requires extra boundary conditions. The question is how to set general boundary conditions for both fields to grant existence to the energy minimizers.

Finally, no hydrodynamic considerations have entered our study, but the question should be asked as to whether the relaxation in time of $\twist$ represents a further source of dissipation.

\section*{Acknowledgements}
I wish to thank Oleg D. Lavrentovich for his encouragement to pursue this study and for his kindness in providing me with some of his results prior to their publication. I am also greatful to Mikhail A. Osipov for having suggested studying the work of Lorman and Mettout \cite{lorman:theory} on the symmetry of helical nematics. I am finally indebted to the kindness of two anonymous Reviewers whose critical remarks and constructive suggestions improved considerably this manuscript.


\end{document}